# Volumetric Additive Manufacturing of Silica Glass with Microscale Computed Axial Lithography

Joseph Toombs[1]*, Manuel Luitz[2], Caitlyn Cook[3], Sophie Jenne[2], Chi Chung Li[1], Bastian Rapp[2,4,5], Frederik Kotz-Helmer[2,4,5], Hayden Taylor[1]*

**Affiliations:**

[1] Department of Mechanical Engineering, University of California, Berkeley; Berkeley, CA 94720, United States of America.

[2] Department of Microsystems Engineering, Albert Ludwig University of Freiburg; 79104 Freiburg, Germany.

[3] Lawrence Livermore National Laboratory; Livermore, CA 94550, United States of America.

[4] Glassomer GmbH; Georges-Köhler-Allee 103, 79110 Freiburg, Germany.

[5] Freiburg Materials Research Center (FMF), Albert Ludwig University of Freiburg; 79104 Freiburg, Germany.

*Corresponding authors. Email: jtoombs@berkeley.edu, hkt@berkeley.edu

**Abstract:** Glass is increasingly desired as a material for manufacturing complex microscopic geometries, from the micro-optics in compact consumer products to microfluidic systems for chemical synthesis and biological analyses. As the size, geometric, surface roughness, and mechanical strength requirements of glass evolve, conventional processing methods are challenged. We introduce microscale computed axial lithography (micro-CAL) of fused silica components, by tomographically illuminating a photopolymer–silica nanocomposite which is then sintered. We fabricated 3D microfluidics with internal diameters of 150 μm, freeform micro-optical elements with surface roughness of 6 nm, and complex high-strength trusses and lattice structures with minimum feature sizes of 50 μm. As a high-speed, layer-free digital light manufacturing process, micro-CAL can process extremely viscous nanocomposites with high geometric freedom, enabling new device structures and applications.

**One-Sentence Summary:** Light-based tomographic 3D printing of silica glass enables fabrication of diverse geometries with low roughness and high transparency.

**Main Text:**

The uses of glass are innumerable because of its optical transparency, thermal and chemical resistance, and low coefficient of thermal expansion. Established applications in architecture, consumer products, optical systems, and art have been joined by specialized uses such as fiber optics in communication, diffractive optics in augmented reality, and lab-on-a-chip devices for chemical and biological analyses (*1–3*). With increased specialization come more demanding requirements for geometry, size, and optical and mechanical properties. Additive manufacturing (AM) has emerged as a promising technique to meet challenging new combinations of requirements. AM of glass materials has been achieved with fused filament fabrication of molten glass (*4, 5*), selective laser melting of pure glass powder (*6, 7*), direct ink writing of silica sol-gel inks (*8*), stereolithography (SLA) (*9, 10*) and multiphoton direct laser writing (DLW) (*11*) of silica nanocomposites which consist of silica nanoparticles dispersed in a photopolymerizable organic liquid.

All these methods employ serial material deposition or conversion which can limit geometric freedom. Layering-induced defects can also affect the printed object's optical and mechanical properties (*5, 12*). In this work, we introduce volumetric AM (VAM) of glass nanocomposites. VAM describes techniques which polymerize whole 3D objects simultaneously in a volume of precursor material, circumventing the need to build objects layer-by-layer. VAM methods based on holographic exposure (*13*), orthogonal superposition (*14*), and tomographic principles (*15, 16*) are enabled by specialized optical engineering and photopolymer synthesis. The tomographic technique of computed axial lithography (CAL) polymerizes 3D structures by the azimuthal superposition of iteratively optimized light projections from temporally multiplexed exposures (Fig. 1A) (*15, 17, 18*). CAL has several advantages for processing glass nanocomposites. There is no relative motion between the precursor material and the fabricated object during printing, so high-viscosity and thixotropic nanocomposite precursors can easily be used. The layer-less nature of the process enables smooth surfaces and complex geometries. Because the fabricated object is surrounded by precursor material during printing, sacrificial solid supporting structures are not needed. These attributes are desirable for applications including micro-optical components and microfluidics.

In this work, we sought production of microscale features, so we constructed a 'micro-CAL' apparatus (Fig. 1B) which coupled a laser light source into an optical fiber with small mode field size and low numerical aperture (*16*) and demagnified the light pattern defined by the digital micromirror device. This design minimized the system's étendue and hence the divergence and blurring of light. We measured optical resolution in terms of the modulation transfer function (MTF) — the level of contrast transfer by the complete optical system as a function of spatial frequency. We achieved MTF greater than 0.4 at frequencies $\geq$ 66.7 cycles/mm in the central 1.5 mm diameter of the build volume (Fig. S2–S4). Combined with gradient descent digital mask optimization (*15*), the micro-CAL system enabled rapid printing (within about 30–90 seconds) of microstructures with minimum feature sizes of 20 μm and 50 μm in polymer and fused silica glass, respectively (Fig. S6, Fig. 3, Fig. 4).

For the fused silica prints, we used a photocurable μSL v2.0 resin with high transparency (Fig. 2A and supplementary text) consisting of a liquid monomeric photocurable binder matrix and solid amorphous spherical silica nanoparticles with nominal diameter of 40 nm (see materials and methods). The binder was polymerized via free radical polymerization and supported the nanoparticles in the printed construct. After printing, structures were removed from the volume of nanocomposite resin and surplus resin was reused for later prints. The structures were developed by rinsing in ethanol or propylene glycol methyl ether acetate to remove excess uncured resin. The resulting green parts were subjected to thermal treatment in two steps: debinding and sintering (Fig. 1A, Table S3, Table S4). The debinding treatment burned out the polymer binder matrix resulting in a porous silica brown part. During sintering the nanoparticles of the brown part fused together forming a dense transparent glass part. Isotropic shrinkage (Fig. S10) occurred during sintering, so it was necessary to scale parts in computer-aided design prior to fabrication to account for the dimensional change (see materials and methods).

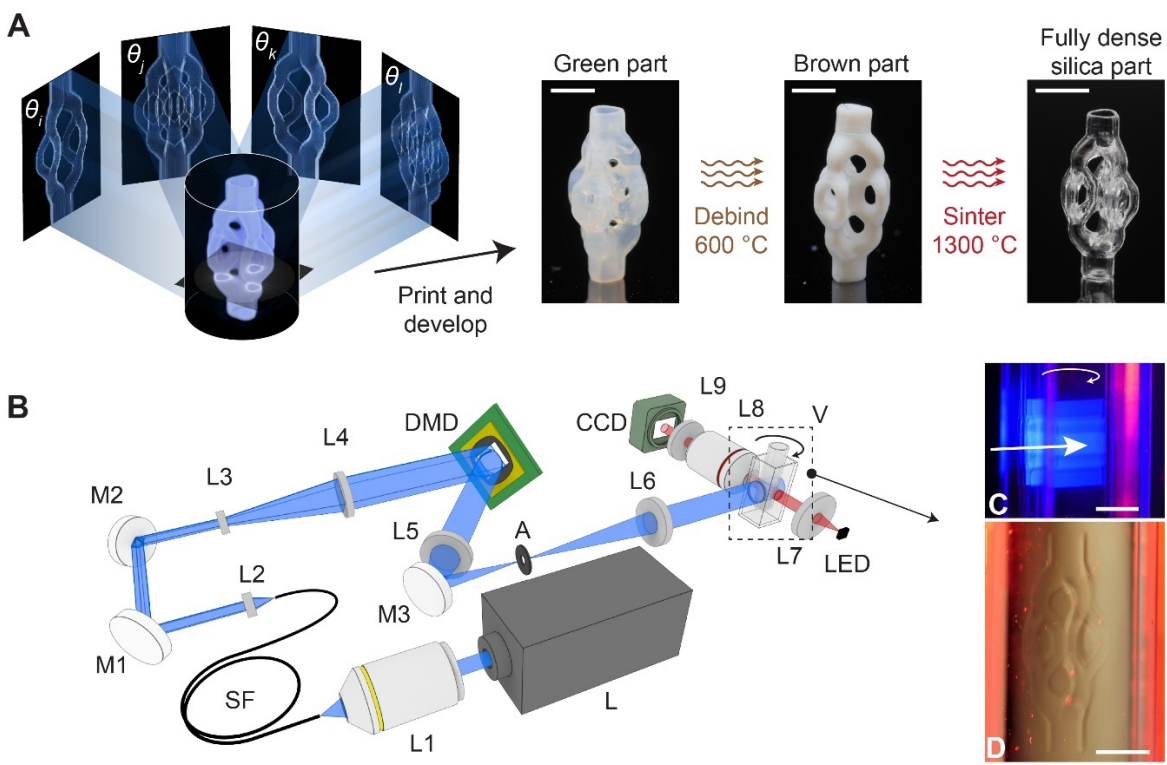

**Fig. 1. Printing transparent fused silica glass with micro-CAL.** (**A**) Tomographic superposition produces a 3D light dose which selectively polymerizes a geometry. After printing, the part is developed by rinsing away residual resin in a solvent. Debinding and sintering steps follow. Scale bars: 2 mm. (**B**) Optomechanical setup. Detailed description of the setup and components are given in the materials and methods. (**C**) The projected image propagates into the rotating vial. Scale bar: 5 mm. (**D**) Immediately after light exposure, the printed object can be observed in the container weakly distorting the background. Scale bar: 2.5 mm.

The tomographic illumination process of CAL means that material that is outside the target geometry receives an appreciable light dose. To achieve selective material conversion, the resin precursor therefore has a threshold light exposure dose below which polymerization is negligible. In the prior VAM research, the induction period—the period of time in which conversion is inhibited by radical scavenger species in the resin—was a result of oxygen inhibition (*14–16*). However, the glass nanocomposite used in this work exhibited a small natural induction period. Besides molecular oxygen, several molecules including quinones and nitroxides, such as 2,2,6,6-tetramethylpiperidinoxyl (TEMPO), are recognized as effective radical inhibitors (*19, 20*). We added various concentrations of TEMPO to the resin, and performed real-time UV Fourier transform infrared spectroscopy (FTIR) analysis to determine the effect of TEMPO concentration on the inhibition time (Fig. 2B and materials and methods). The addition of TEMPO increased the duration of the induction period and had negligible effect on the kinetics of polymerization and maximum degree of conversion. The sharply nonlinear relationship between conversion and exposure dose provided by TEMPO significantly improved the selectivity of the micro-CAL process. Fig. 2D and 2E show an example of such improvement using the same set of digital light projections. With TEMPO, conversion inside the void of the cubic cage was reduced, and removal of uncured material was more easily achieved than in the nanocomposite without TEMPO. This improvement enabled fabrication of diverse geometries with positive feature sizes as low as 50 μm in the μSL v2.0 material (Fig. 3). In pure monomeric resins, we achieved significantly smaller positive feature sizes as low as 20 μm (Fig. S6 and S7). We attribute this resolution enhancement for polymeric structures to the absence of solid nanoparticles which results in less light scattering (see supplementary text for theoretical scattering of μSL v2.0) and easier development due to lower resin viscosity and stronger green state.

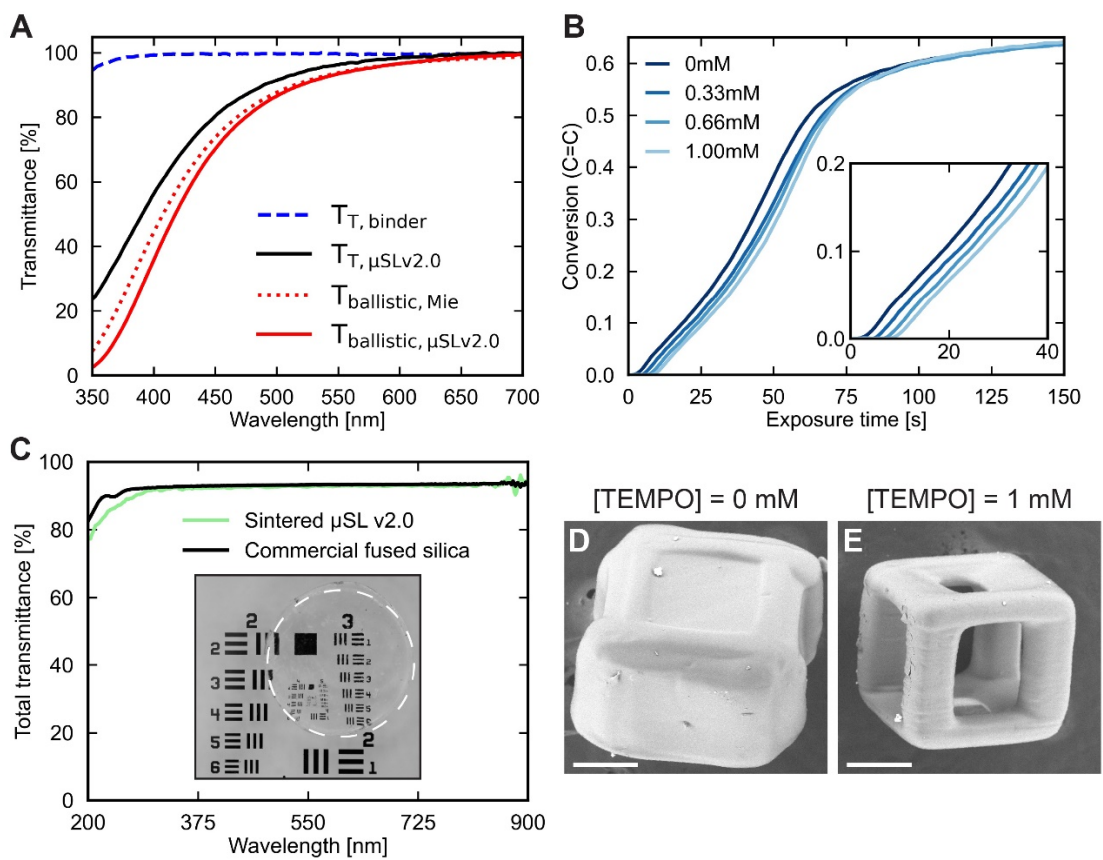

**Fig. 2. Resin and sintered silica material characterization.** (**A**) Transmittance spectra of binder (resin without silica), $T_{T,binder}$, and μSL v2.0 (with silica) where $T_{T, \mu SLv2.0}$ is total transmittance, $T_{ballistic, \mu SLv2.0}$ is the ballistic (collimated) component, and $T_{ballistic, Mie}$ is estimated by Mie theory. (**B**) UV FTIR measurement of carbon–carbon double-bond conversion as a function of exposure time and concentration of TEMPO. Increased induction period is observed with increased TEMPO concentration. (**C**) Total transmittance spectra of micro-CAL-printed and sintered μSL v2.0 disk and commercial fused silica coverslip. Inset image shows qualitatively the optical transparency of a printed circular disk (inside dashed circle) in front of a USAF resolution target. See materials and methods for experimental details.

(**D**, **E**) SEM micrographs of cube cages printed and developed with 0 mM and 1 mM [TEMPO], respectively. Scale bars: 200 μm.

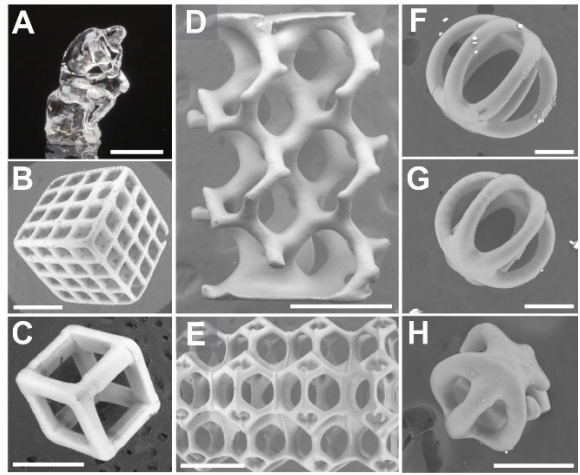

**Fig. 3. CAL-printed glass structures.** (A) Rodin's 'Thinker', (B,C) cubic lattice structures, (D) skeletal gyroid lattice with minimum positive feature size of 50 μm, (E) tetrakaidecahedron lattice, (F–H) spherical cage structure with minimum positive feature sizes of 75 μm, 60 μm, and 50 μm, respectively. (A: photograph; B–H: scanning electron microscope (SEM) micrographs). Scale bars: (A–E) 1 mm; (F–H) 200 μm.

Synthetic microstructured cellular materials have found use in a variety of fields including photonics, energy, bioengineering, and desalination, as well as in high-temperature environments (*21–23*). Specifically, mechanical metamaterials, designed to exhibit mechanical properties which are unattainable by the bulk material, e.g. negative Poisson ratio, are emerging as an important area in AM because they often have a porous nature that is challenging to reproduce with conventional manufacturing techniques (*24*). In contrast with SLA, DLW, and fused filament fabrication, CAL builds objects volumetrically, which means complex, low-relative-density lattice and truss structures can be created in any orientation without supporting material (Fig. 3B–D, Fig. 4A). We fabricated tetrakaidecahedron lattices from transparent fused silica glass with strut elements about 100 μm in diameter (Fig. 3E). For specific applications in which orientation of the microstructure is critical, volumetric processing may prove useful because it eliminates defects due to layering that would be present in certain print orientations using other AM techniques.

We measured low mean $R_a$ roughness of 0.407 μm on several members of trusses (as in Fig. 4B) using laser confocal profilometry (Table S6). To demonstrate the mechanical properties of a micro-CAL-printed object, we fabricated a Howe truss (*25*) and subjected it to three-point bend loading (Fig. 4B and C, Fig. S11). The nominal tensile stress at failure was 187.7 MPa, which is higher than the fracture strength of approximately 100 MPa that is typically measured in silica glass components fabricated by conventional means, e.g., casting or drawing (*26*). VAM may limit the creation of microcracks and indentations which would otherwise compromise fracture strength. Micro-CAL could be used to investigate novel high strength lattices which exploit silica's high intrinsic strength and strain at failure in the absence of large flaws (*27*).

Fused silica glass microfluidic devices offer many advantages over polymeric devices including high resistance to temperature and harsh acids and organic solvents, as well as high optical transmission over an extended UV, visible, and infrared range. However, conventional fabrication techniques such as planar lithographic processes require toxic fluoric etchants and are largely limited to 2D (*28*). With micro-CAL, we achieved rapid freeform fabrication of perfusable branched 3D microfluidics with low surface roughness, high transparency, and channel diameters and wall thickness as low as 150 μm and 85 μm, respectively (Fig. 4D, 4E, and Fig. S13). These properties show that micro-CAL has the potential to advance the fabrication of microreactors which are important for parallel drug screening and highly controlled flow synthesis (*3*).

The demand for more compact, lightweight, and high-quality cameras in consumer electronics and biomedical imaging pushes development of advanced millimeter-scale optical systems. AM has enabled production of freeform refractive microlenses designed for specific applications, e.g. foveated imaging. However, imaging elements made by layer-based techniques require post-processing such as polishing or coating to suppress the scattering induced by layer artifacts (*29*). With CAL, the 3D light dose possesses a radially and axially oriented gradient which has the effect of smoothening optical surfaces. We measured $R_a$ roughness as low as 6 nm on as-fabricated lens surfaces using laser scanning confocal microscopy (Fig. S17). We demonstrated printing of several refractive optical elements including an air-spaced doublet aspheric lens optimized for operation at 532 nm wavelength, hexagonal and lenticular microlens arrays, and a spherical Fresnel lens (Fig. 4F–I). The full-width, half-maximum of the point spread function (PSF) under collimated 532 nm illumination is less than 50 μm for each element. Low figure error in the range of 1–10 μm was achieved for spherical surfaces; however, figure error up to 60 μm persists for the aspheric lens surface and imaging remains a challenge (Fig. S15 and S16). These results suggest that while roughness is on par with commercial optics, figure error should be improved, perhaps by utilizing in-situ feedback and correction algorithms during printing (*16*, *30*).

The micro-CAL system we have developed enables manufacturing of structures with minimum feature sizes of 20 μm in polymer and 50 μm in fused silica, with unprecedented geometric freedom, low surface roughness, and high fracture strength and optical transparency in fused silica. Through optical engineering and specialized photopolymer development, we have established a glass fabrication framework that merges the facile processing of silica nanocomposites with layer-less volumetric additive manufacturing, and could advance research in and industrial application of mechanical metamaterials, 3D microfluidics, and freeform optics.

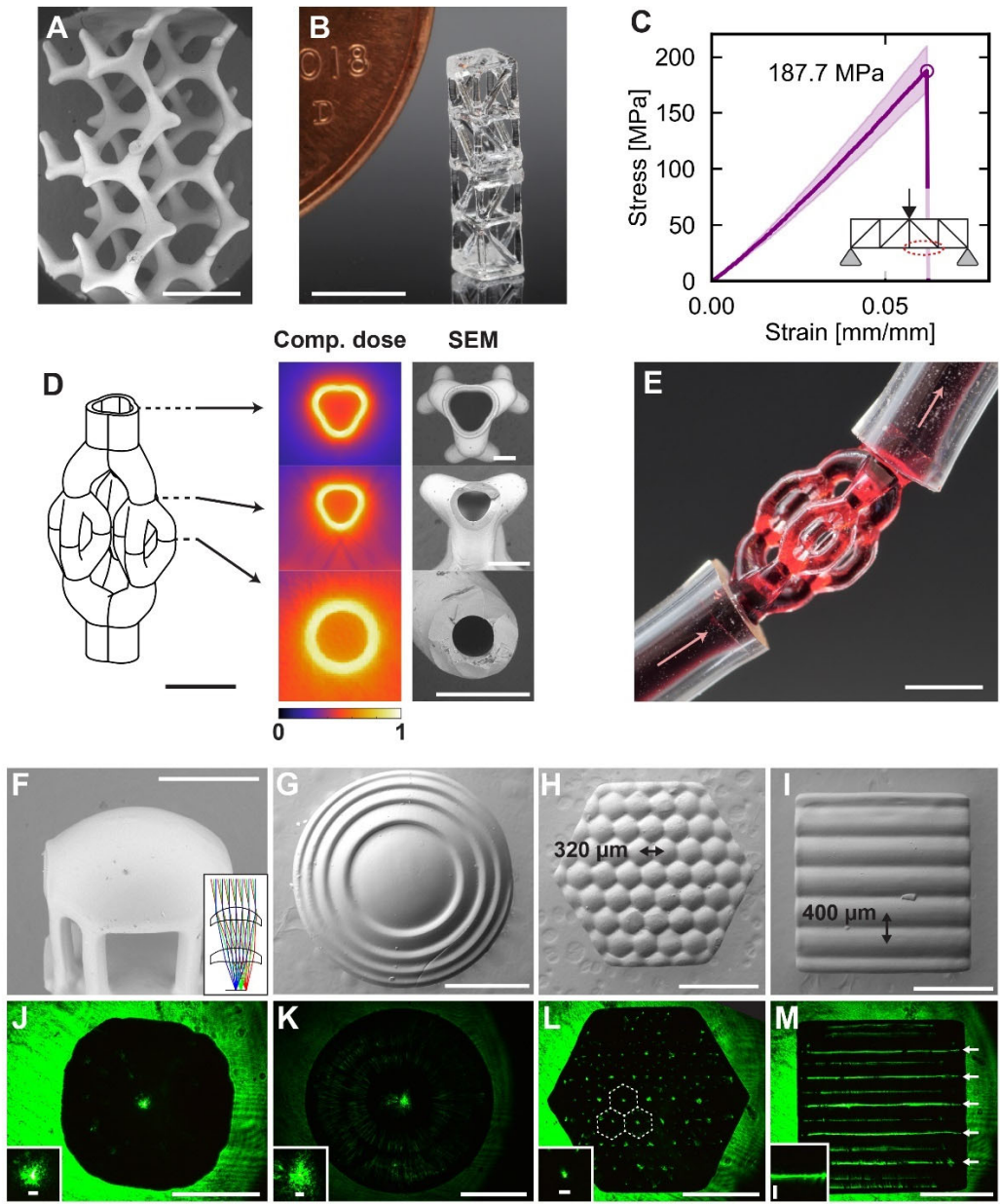

**Fig. 4. Applications of glass CAL-printed microstructures.** (**A**) SEM micrograph of skeletal gyroid lattice. Scale bar: 1 mm. (**B**) Photograph of Howe truss. Scale bar: 2 mm. (**C**) 3-point bend test loading of truss structure where stress is the tensile stress in the bottom member indicated in the schematic. The filled region represents the bounded range of possible stresses due to variation in diameter of members. (**D**) Schematic of trifurcated channel with normalized computational dose profiles and SEM cross sections at representative slices along a channel revealing three levels of pore sizes, 750 µm, 350 µm, and 215 µm (top to bottom). Scale bar: (schematic) 2mm, (SEM micrographs) 500 µm. (**E**) Dyed liquid passed through the model demonstrates perfusability. Scale bar: 2 mm. (**F–I**) SEM micrographs of printed optical elements. (**J–M**) PSF's of optical elements in F–I following focusing of 532 nm laser illumination. Insets show zoomed PSF. Scale bars (F–M): 1 mm (insets J–M: 50 µm)

**Acknowledgments:** We thank the staff at the University of California Berkeley Electron Microscopy Lab for advice in electron microscopy sample preparation and data collection.

**Funding:**

    National Science Foundation under Cooperative Agreement No. EEC-1160494 (JT, CL, HT)

    European Research Council (ERC) under the European Union's Horizon 2020 research and innovation program (grant agreement No. 816006) (ML)

    European Research Council (ERC) under the European Union's Horizon 2020 research and innovation program (grant agreement No. 825521) (SJ)

    German Research Foundation (Deutsche Forschungsgemeinschaft, DFG) project number 455798326 (FK)

    Lawrence Livermore National Laboratory Directed Research and Development program. The work was performed under the auspices of the U.S. Department of Energy by Lawrence Livermore National Laboratory under contract DE-AC52-07NA27344 (LLNL-JRNL-826682). (CC)

**Author contributions:**

    Conceptualization: JT, FK, HT

    Methodology: JT, ML, FK

    Investigation: JT, ML, CC, SJ, CL

    Visualization: JT

    Funding acquisition: HT, FK, BR

    Project administration: HT, FK, JT

    Supervision: FK, HT, BR

    Writing – original draft: JT

    Writing – review & editing: All authors

**Competing interests:** HT has U.S. patent 10,647,061 B2 relating to computed axial lithography. BR and FK have U.S. patent 10,954,155 B2 relating to the silica nanocomposite material described in this paper. FK and BR are co-founders of, and have an equity interest in, Glassomer GmbH. Authors declare that they have no other competing interests.

**Data and materials availability:** All data are available in the main text or the supplementary materials. Code used to analyze and visualize data is available upon request.

# Supplementary materials for

## Volumetric Additive Manufacturing of Silica Glass with Microscale Computed Axial Lithography

Joseph Toombs, Manuel Luitz, Caitlyn Cook, Sophie Jenne, Chi Chung Li, Bastian Rapp, Frederik Kotz-Helmer, Hayden Taylor

Correspondence to: jtoombs@berkeley.edu, hkt@berkeley.edu

**This PDF file includes:**

Materials and Methods
Supplementary Text
Figs. S1 to S19
Tables S1 to S6

## Materials and Methods

Optomechanical setup

Prior CAL printers have used LED illumination with large optical étendue — the product of the light source size and divergence angle. A consequence of conservation of étendue is that divergence angle increases if the source is demagnified. In CAL, large divergence would cause blurring of the optical dose, and, consequently, printed features. The optomechanical setup used in this work minimizes étendue by using a laser light source (*16*) in the configuration shown in Fig. 1B and described below.

Lens 1 (L1) launches the 442 nm continuous wave laser (Fig. S1) into the multimode square core fiber (SF). The SF provides a homogeneous intensity in the shape of a square which more efficiently fills the rectangular face of the digital micromirror device (DMD) than a circular Gaussian profile (*16*, *31*). L2 collimates the output of the fiber. Mirrors 1 and 2 (M1, M2) align the beam to the entrance of the beam expander (L3, L4). The expanded square beam profile reflects from "on"-state pixels on the DMD into the subsequent optics, whereas light reflected from "off"-state pixels is directed into a beam dump. The DMD is rotated 45° such that the tilt axis of each micromirror is vertical. The DMD is imaged into the print container (V) with the 4f system created by L5 and L6. The level of demagnification is set to either 0.45× or 1.25× by the selection of the lens for L6. An aperture (A) blocks unwanted diffraction orders from the DMD. The LED, L7 – L9, and the CCD camera form a simple shadowgraph system for in-situ imaging.

Resolution and modulation transfer function

The theoretical demagnified pixel size at the focal plane of the projection optical system is 4.9 µm and 13.5 µm in the 0.45× and 1.25× demagnification configurations, respectively (assuming a 10.8 µm pixel size on the DMD). Equivalently, the theoretical spatial frequencies are 102 cycles/mm and 37 cycles/mm, respectively. The optical system should have a modulation transfer function (MTF) capable of passing such frequencies, ideally over the entire build volume. However, due to conservation of étendue this is challenging. The fiber core size is 50 × 50 µm$^2$ with numerical aperture NA = 0.2. After beam expansion to illuminate the DMD, the square image is 12 × 12 mm$^2$ with NA = 0.00083. This corresponds to NA = 0.0019 and NA = 0.00067 after demagnification in the 0.45× and 1.25× configurations, respectively. At 1 mm radius for 0.45×, the theoretical pixel size expands to 7.5 µm (66.7 cycles/mm) assuming the refractive index of the resin is 1.474.

The MTF of the projection optical system was measured using a method similar to the slanted-edge MTF measurement which is used in fields from microscopic to satellite imagery (*32*). However, in this case, the target "slanted edge" is loaded as a pattern on the DMD and the camera sensor images the projected pattern in the print volume. The "slanted edge" is not slanted on the DMD; it is simply the left 960×1080 pixels in the "on" state and the right 960×1080 pixels in the "off" state and the camera sensor is rotated such that the edge appears at a 6–10° angle from vertical in the image (Fig. S2B). In this configuration, the measured MTF of the optical system incorporates the effect of imperfect dark–light contrast ratio of the DMD due to diffraction efficiency and fill factor being less than 100%. The MTF is calculated by first projecting pixel intensities onto a vector perpendicular to the slanted edge (Fig. S2A). This has the effect of supersampling the edge at a sampling interval smaller than the sensor pixel pitch. The sensor pitch is 598 pixels/mm (1.67 µm pixel size). The sampled points are then smoothed with a mean filter to obtain the edge spread function (ESF) (Fig. S2C). The ESF is then resampled on a uniform interval and differentiated to obtain the line spread function (LSF). A Gaussian function is fit to the LSF, and the Fourier transform is taken to obtain the MTF (Fig. S2D).

The MTF was measured for several defocus positions within the build volume. Shown in Fig. S4 is the MTF within the central 1.5 mm of build volume in the 0.45× configuration. Spatial frequencies ≤ 100 cycles/mm are readily transferred with MTF ≳ 0.45 near the focus, but farther away from the focus only spatial frequencies ≤ 60 cycles/mm are transferred at this contrast. The optical system realizes theoretical resolution near the focus, but due to conservation of étendue the resolution diminishes away from the focus. The measurements agree with the theoretical resolution albeit with slightly lower MTF, which is expected to arise from practical limitations like minor misalignment and variation in the actual fiber core size and NA.

Preparation of silica nanocomposite resin

Silica glass nanocomposite resin (Glassomer μSL v2.0) with a solid loading of 35 vol% silica was supplied by Glassomer GmbH. The nanocomposite's liquid monomer binder consisted of a mixture of trimethylolpropane ethoxylate triacrylate (TMPETA) and hydroxyethylmethacrylate (9). To the resin was added 0.117 wt% camphorquinone (CQ) and equal mass ethyl 4-(dimethylamino)benzoate (EDAB) to make a 0.01 M photoinitiator concentration. A 0.2 M solution of 2,2,6,6-tetramethyl-1-piperidinyloxy (TEMPO) photoinihibitor in TMPETA was created. To the resin was added a volume of the TEMPO-TMPETA solution to create a 0.001 M concentration of TEMPO (0.5 vol%) in the nanocomposite resin. CQ, EDAB, TMPETA, and TEMPO were purchased from Sigma-Aldrich.

Experimental 3D printing

Prepared resin was added to 1 mL glass shell vials (ThermoFisher Scientific). The vial was loaded into a pair of variable aperture irises mounted on the rotation stage (PRM1Z8, Thorlabs). Projection images were displayed on the DMD and were synchronized to the rotation of the vial. The print was terminated when deemed complete by visual inspection through the in-situ shadowgraph imaging system (Fig. S5). After printing, the part was rinsed in ethanol or propylene glycol methyl ether acetate (PGMEA) and slowly agitated in a vortex mixer. For finer features and microfluidic channels, the resin-solvent mixture was heated to 85 °C to reduce its viscosity and promote dissolution of the resin.

Minimum feature size in polymeric materials

The resin used to fabricate polymeric structures consisted of pentaerythritol tetraacrylate monomer (TCI America), 0.01 M CQ, and 0.001 M TEMPO. After structures were printed they were rinsed and developed in PGMEA. After development was complete, contaminated PGMEA was exchanged with fresh PGMEA without removing the structures from the liquid. A small amount of Irgacure 369 photoinitiator was added to the solvent and allowed to dissolve. The structures residing in this solution were flood exposed with 365 nm light (Thorlabs M365LP1) for 5 minutes. The structures then had sufficient strength to avoid collapse upon removal from the liquid. Fig. S6 and Fig. S7 show periodic lattice structures printed in polymer. The minimum feature size obtained was 20 μm in either lattice.

Refractive index of silica nanocomposite resin

The refractive index of each uncured samples of pure monomer binder and µSL v2.0 were measured with an Anton Paar Abbemat mulitiwavelength refractometer (Fig. S8). All measurements were performed at 20 °C. Cauchy's equation was fit to measurements made at eight points. Note that all measurements were performed using uncured liquid samples. Since we terminated light exposure immediately after the resin gelation point which occurs at low carbon–carbon double-bond conversion for multifunctional acrylate systems (*33*), significant change in refractive index in the binder material was avoided and was assumed to have a negligible effect on the printing process.

UV Fourier transform infrared spectroscopy
	In order to characterize the increasing induction period with increasing TEMPO concentration, UV curing kinetics were monitored with a Bruker 80 Fourier Transform Infrared spectrometer equipped with a horizontal UV exposure attachment. Resin samples were placed between two glass plates and spacing was kept constant at 0.5 mm via silicon gasket material. An Omnicure S2000 lamp with Thorlabs 441.6 nm 10 nm bandpass filter was used to expose samples, while light intensity was verified prior to each experiment with a Thorlabs power meter and S120VC sensor. Ultraviolet/visible spectroscopy (Thorlabs CCD spectrometer) was performed on the 441.6 nm filtered broadband light source to confirm a 10 nm bandpass (Fig. S9). Conversion was calculated from the change in the acrylate peak area with light exposure in the near-infrared range at about 6165 $cm^{-1}$ and by using the initial unreacted resin peak as a reference. Data was acquired at 5 scans per second and a 4 $cm^{-1}$ resolution (*35–37*).

Resin and sintered fused silica optical characterization
	Resin optical characterization was performed on a Shimadzu UV-3600i Plus with integrating sphere attachment. All measurements were performed in cuvettes with 1 cm pathlength. Sintered glass optical characterization was performed on a Shimadzu UV-2600 Plus with integrating sphere attachment. The commercial fused silica coverslip was 250 µm thick and the micro-CAL-printed disk was 290 µm thick.

Heat treatment
	Debinding of the printed green parts was completed in an ashing furnace (11/3 AAF Carbolite-Gero, Germany). Sintering of the brown parts was completed in a high temperature dental furnace (AUSTROMAT 664iSiC, Dekema, Germany). Table S3 and Table S4 provide the debinding and sintering thermal profiles, respectively.

Shrinkage
	To verify that the tomographic light dose superposition has no effect on the isotropy of shrinkage, thin disks were printed in different orientations and their diameters measured before and after thermal treatment (Fig. S10). Linear shrinkage of 26 % was measured irrespective of printed orientation.

Mechanical characterization
	Mechanical strength testing was performed on a Mark 10 ESM1500 with MR03-10 force sensor and Model 5i force gauge. Printed trusses were loaded at 1.5 mm/min in a 3-point bending configuration in which cylindrical supports were placed underneath the sample. Similarly, to the upper compression plate was attached a cylindrical "anvil" (Fig. S11).

During loading in section C, the truss experiences linear elastic deformation with one minor failure. At the end of section C, major (but not complete) failure occurs in the bottom member as indicated in Fig. S11D and suggested by the brittle stress-strain failure behavior in Fig. S11F. The truss tested in Fig. 4B (which is different from the truss referred to here) exhibited complete failure at this point. In section D, the truss is loaded further until complete failure occurs in the top members as indicated in Fig. S11E.

We estimate the stress at failure in the bottom member by first assuming the members are under pure axial stresses because the members are slender. The free-body diagram of the truss is provided in Fig. S12. In static equilibrium, the net *x*- and *y*-forces and net moment about axis *c* are

$$\sum F_x = -F_2 - F_3 - F_1 \sin \alpha = 0$$

$$\sum F_y = F_1 \cos \alpha + \frac{F_A}{2} = 0$$

$$\sum M_c = \frac{F_A p}{2} + F_3 h = 0$$

where $F_A$ is the applied force. We assume that each member has a circular cross section, therefore, the stress in the bottom member is derived as

$$\sigma_2 = \frac{\frac{F_A}{2}\left(\frac{p}{h} + \tan \alpha\right)}{\pi r^2}$$

where $r$ is the radius of the member.

Refractive lens surface characterization

Spherical lenses were designed with a radius of curvature of 1.6 mm on the convex surface to measure the lens figure error in the sintered state (Fig. S14). The traces in Fig. S15 show three profiles measured over equators of three independent lenses. The mean figure error was less than 10 μm on each lens. Additionally, the top convex surface of the aspheric lens shown in Fig. 4F in the main text was measured with laser scanning profilometry in the brown state (Fig. S16). The mean figure error was larger, in this case, up to about 60 μm. Roughness was also measured on this surface (Fig. S17). Arithmetic mean roughness as low as 6 nm was measured.

**Supplementary Text**

Optical scattering model

To realize high resin transparency, a binder with refractive index close to that of silica was chosen (Fig. S8). We used Mie theory (*38, 39*) to estimate the ballistic transmittance T$_{ballistic,Mie}$ of the nanocomposite given the refractive index contrast and particle size (Fig. 2A) and verify that the optical penetration depth is sufficiently high at the illumination wavelength, of 442 nm, for effective tomographic exposure. The Python package *miepython* was used to perform the Mie theory computations (*40, 41*).

First, the relative refractive index $m'$ is calculated as

$$m'(\lambda) = \frac{n_{silica}(\lambda)}{n_{binder\,matrix}(\lambda)}$$

where $n_{silica}(\lambda)$ is the refractive index of the silica nanoparticles (obtained from the Sellmeier formula for fused silica (Fig. S8)) and $n_{binder\,matrix}(\lambda)$ is the measured refractive index of the monomer excluding the silica nanoparticles. The refractive index of silica is assumed to have no imaginary part since silica has no absorption in the visible spectrum of interest. The wavelengths for which the computations are performed are modified to account for the binder as

$$\lambda' = \frac{\lambda}{n_{binder\,matrix}(\lambda)}.$$

The relative size parameter $x'$, a dimensionless parameter relating the particle circumference to the wavelength, is calculated as

$$x'(\lambda) = \frac{2\pi a}{\lambda'}$$

where $a$ is the particle radius. In this analysis it was assumed that the particle distribution was monodisperse but the calculation can be extended to polydisperse systems (*39*). With $x'(\lambda)$ and $m'(\lambda)$, the scattering phase function and scattering efficiency of the nanocomposite were calculated, respectively, as

$$\frac{I}{I_0} = \frac{\frac{1}{2}(|S_1(\theta)|^2 + |S_2(\theta)|^2)}{k^2 r^2}$$

$$q_{sca}(\lambda) = \frac{4}{x'(\lambda)^2}\text{Re}(S_1(0))$$

where $S_1(\theta)$ and $S_2(\theta)$ are complex amplitude functions of scattering angle, $r$ is distance from the center of the particle, and $k = \frac{2\pi}{\lambda'}$ is the wavenumber (*38*). The absorption efficiency of the nanocomposite is zero because $m'(\lambda)$ is real. The scattering cross section was calculated as

$$\sigma_{sca}(\lambda) = q_{sca}(\lambda)\pi r^2$$

where $q_{sca}$ is the scattering efficiency. The scattering coefficient $\mu_{sca}$ which describes a beam's intensity loss due to scattering in the material is given by

$$\mu_{sca}(\lambda) = \sigma_{sca}(\lambda)N$$

where $N$ is the number density of particles in the material and is defined as

$$N = \frac{\phi}{\frac{4}{3}\pi r^3}$$

where $\phi$ is the volume fraction of particles. $\sigma_{sca}(\lambda)$ and $\mu_{sca}(\lambda)$ are plotted in Fig. S18 for the µSL v2.0 nanocomposite resin. The portion of attenuation of a collimated beam's intensity attributed to scattering is determined by the differential equation

$$-\frac{dI}{dz} = \mu_{sca}I.$$

This leads to a simple exponentially decaying transmission with distance $z$ (*39*)

$$T = \frac{I}{I_0} = exp(-\mu_{sca}z).$$

The simulated transmission as a function of wavelength is plotted in Fig. 2A in the main text. The estimate was validated by UV-VIS spectroscopy equipped with an integrating sphere detector to measure total transmission (scattered and ballistic) and with a conventional detector and aperture to measure ballistic transmission. $T_{ballistic, µSLv2.0}$ is in close agreement with the estimated $T_{ballistic,Mie}$. Although, the theoretical attenuation due to scattering at 442 nm ($\mu_{sca}$ = 0.35 cm$^{-1}$) through the 6 mm inner diameter vial of resin is 19% (Fig. S19A), we neglected scattering to reduce complexity of iterative optimization of projection light patterns. The fidelity of the structures we fabricated suggests that this is a reasonable assumption. Additionally, although the scattering phase function is only somewhat anisotropic (Fig. S19B), the collimated transmission is still high because the scattering efficiency is very small due to small $(m' - 1)$ and small $x'$. While not directly applied to computation of projections in this work, in future investigation, the Mie theory approach could prove useful for developing more accurate models of light propagation in resins having stronger scattering due to larger index mismatch (*16*, *42*).

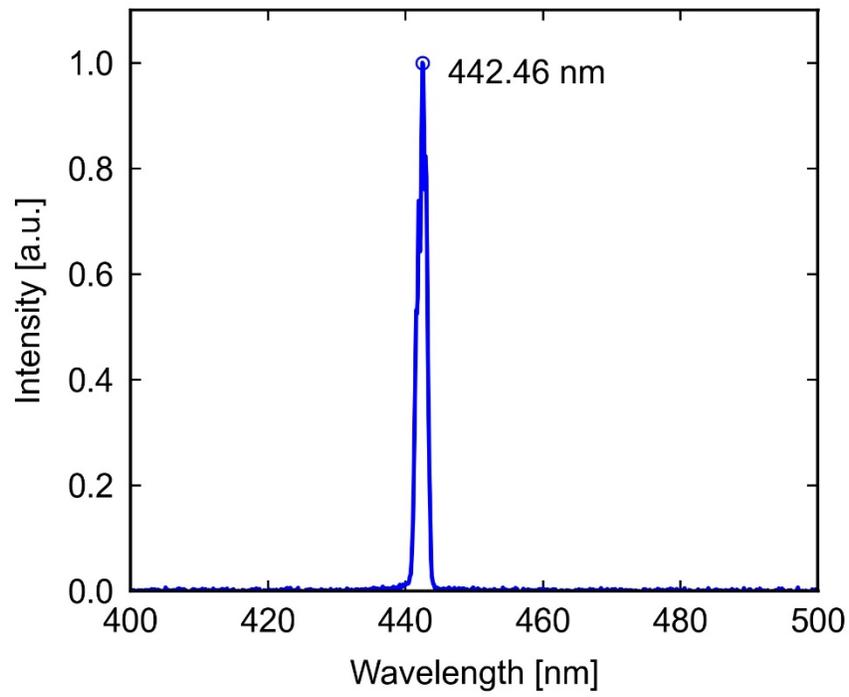

**Fig. S1. Spectrum of the writing laser.** The spectrum was measured with a fiber-coupled Thorlabs CCD spectrometer (CCS100).

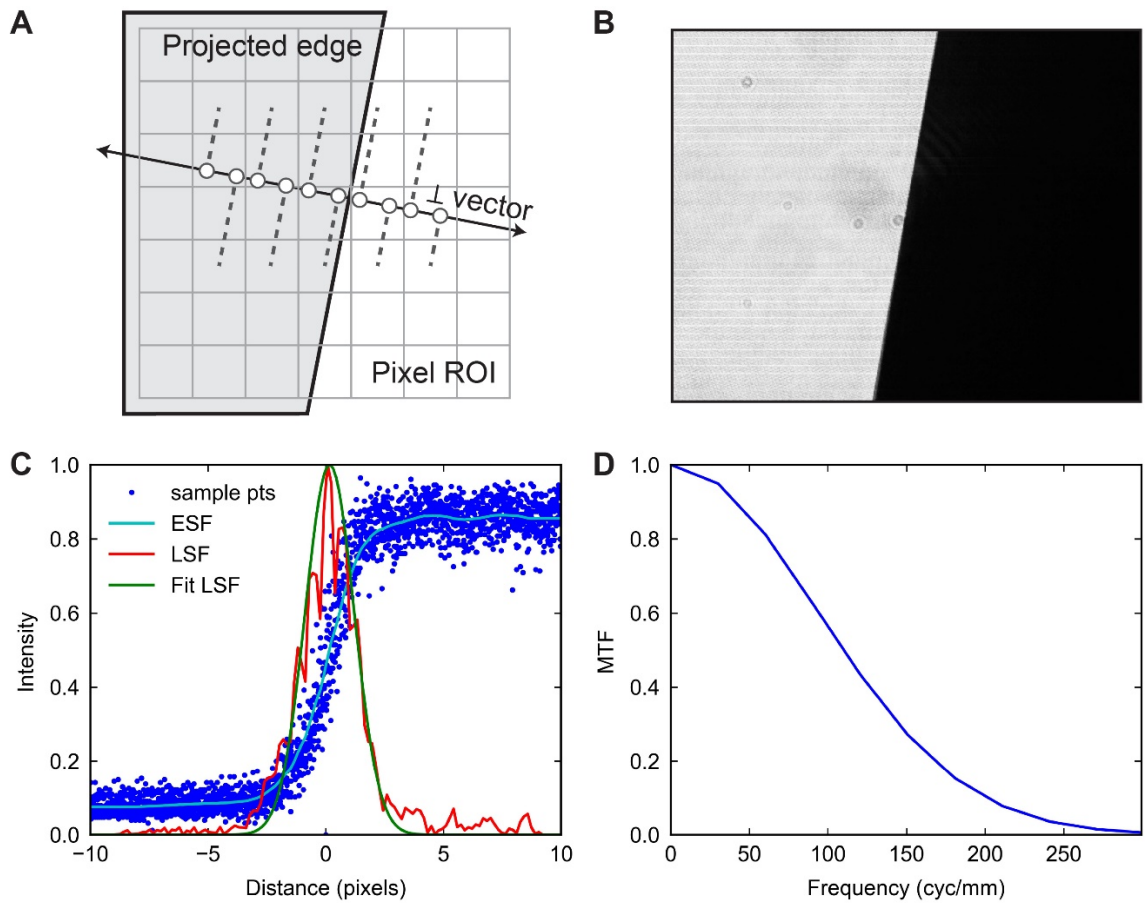

**Fig. S2. Projection optical system MTF calculation.** (**A**) The slanted edge is supersampled by projecting pixel intensities onto a vector perpendicular to the edge. (**B**) Image of the projected edge from a CMOS sensor placed in the print volume. (**C**) Sampled edge, ESF, LSF, and Gaussian fit LSF. (**D**) The Fourier transform of the fit LSF gives the MTF.

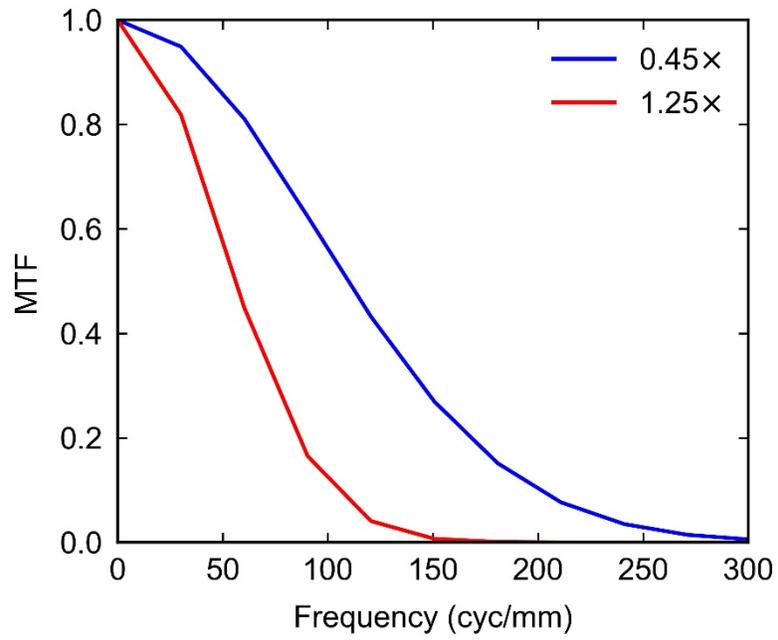

**Fig. S3. MTF projection optical system.** MTF of optical system with DMD demagnification factors of 0.45× (blue) and 1.25× (red). The sensor was placed at the plane of best focus.

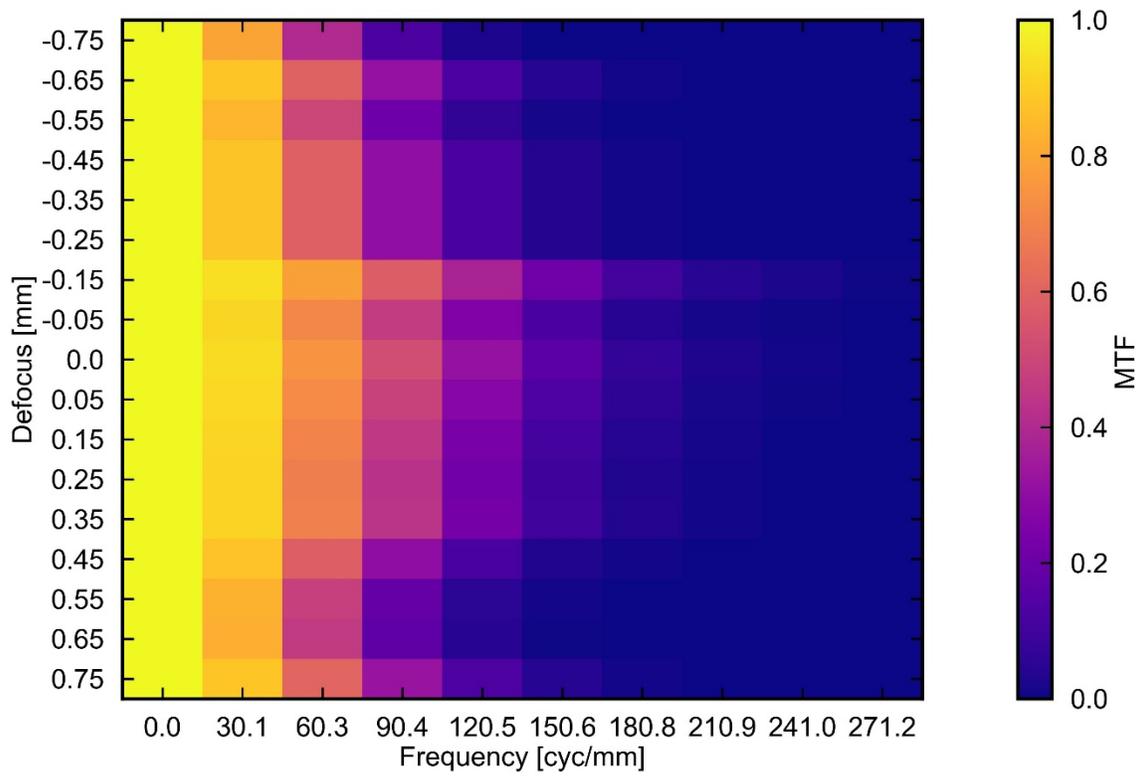

**Fig. S4. MTF of projection optical system in 0.45× demagnification configuration.**

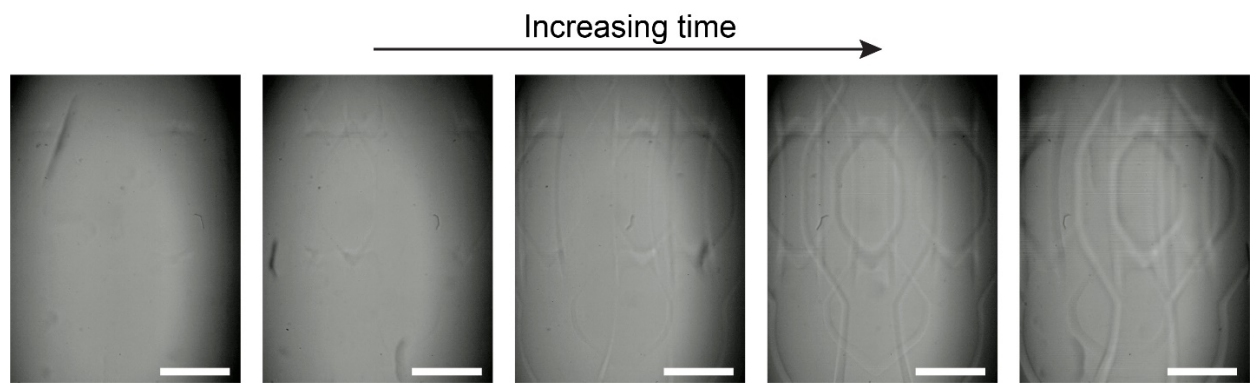

**Fig. S5. In situ shadowgrams during a print.** As exposure time increased, polymerization led to densification of the monomer and consequently an increase in refractive index. The change in refractive index induced small variations in brightness of the shadowgram where the part had formed. Scale bars: 1 mm.

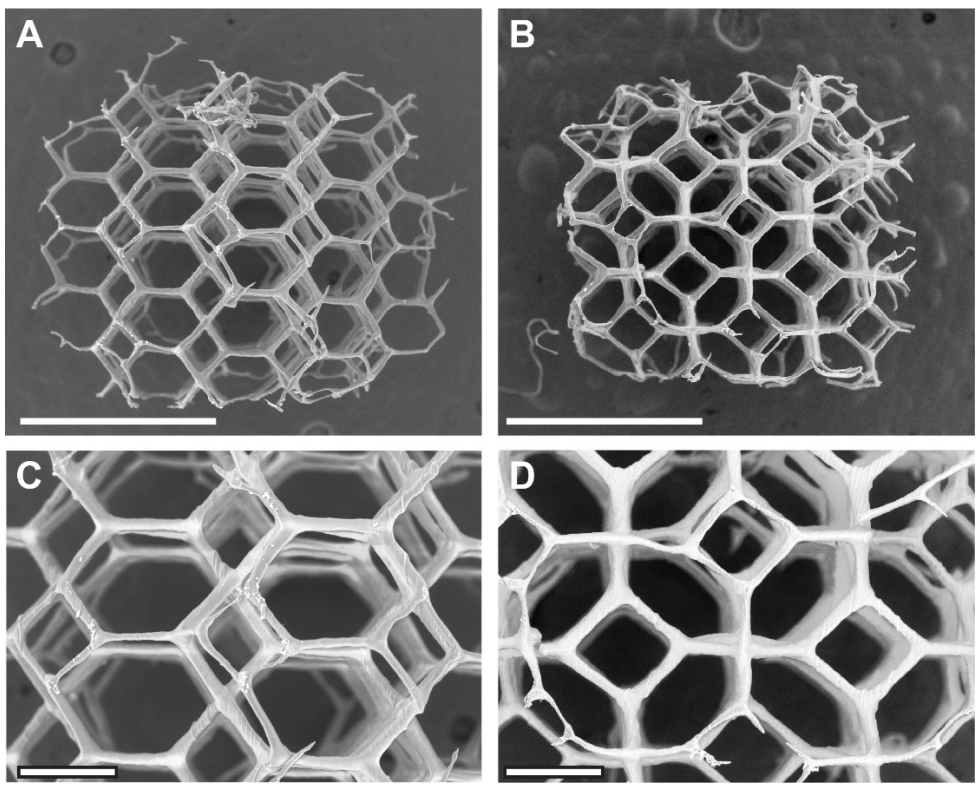

**Fig. S6. Tetrakaidecahedron lattice printed in monomeric resin.** (**A, B**) SEM micrographs of several unit cells of tetrakaidecahedron lattice. Scale bars: 1 mm. (**C, D**) Zoomed SEM micrographs of A and B. Scale bars: 100 μm.

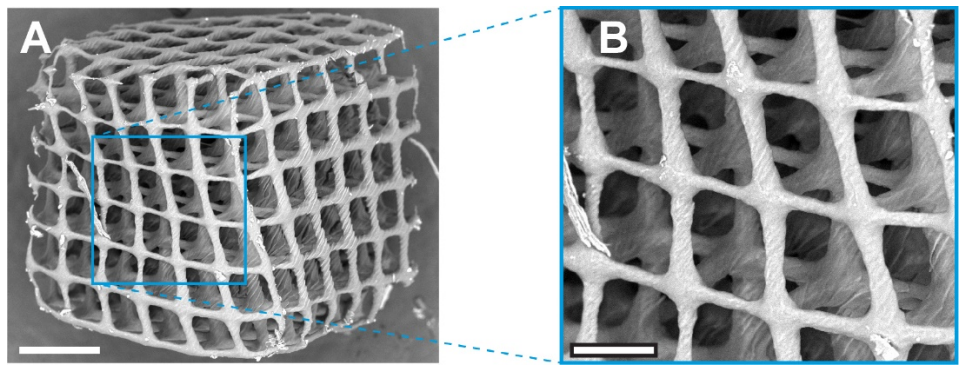

**Fig. S7. Cubic lattice printed in monomeric resin.** (**A**) SEM micrograph of lattice of 7×7×7 cubic unit cells. Scale bar: 500 μm. (**B**) Zoomed SEM micrograph of A. Scale bar: 100 μm.

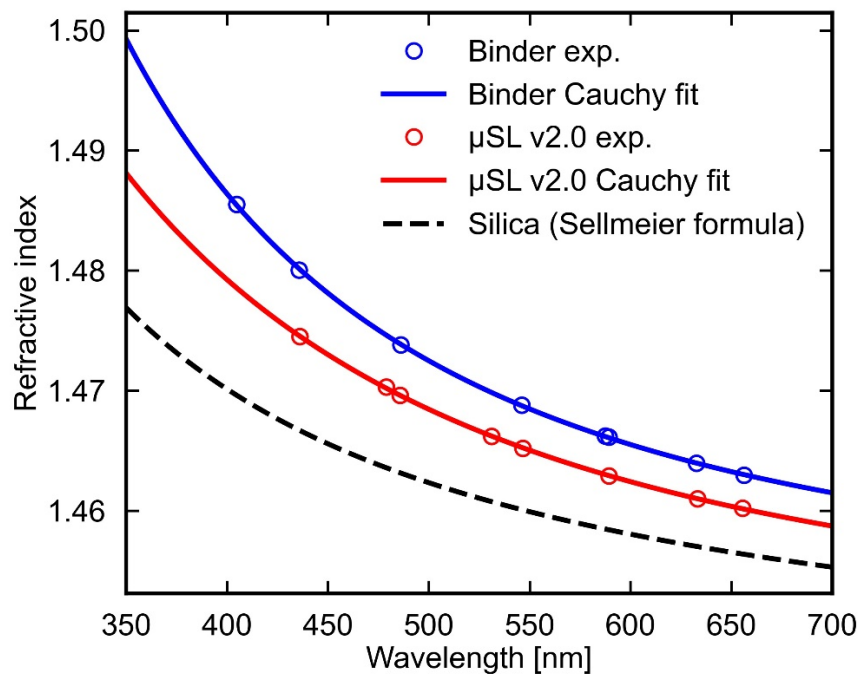

**Fig. S8. Refractive index of µSL v2.0 resin.** The refractive index of the binder material excluding the Aerosol OX50 silica nanoparticles (blue) and the refractive index of the binder including the silica nanoparticles (red) vs. wavelength. Experimentally measured data are shown as open circles with the least squares Cauchy fit in solid lines. The Aerosil OX50 silica nanoparticles refractive index is modeled by the Sellmeier formula for fused silica (*34*). Cauchy fit parameters are given in Table S2.

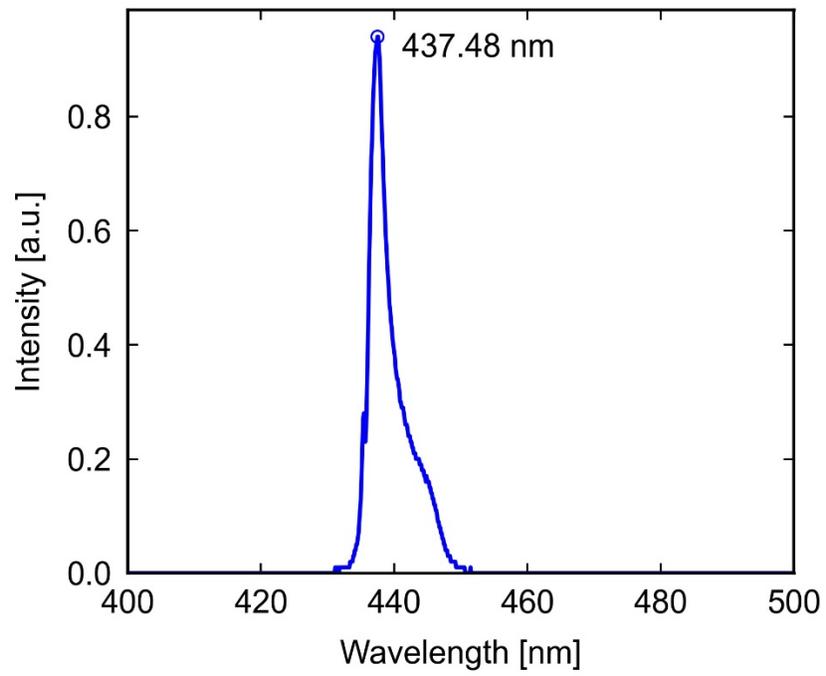

**Fig. S9. Spectrum of Omnicure S2000 with 441.6 nm bandpass filter used in FTIR analysis.**

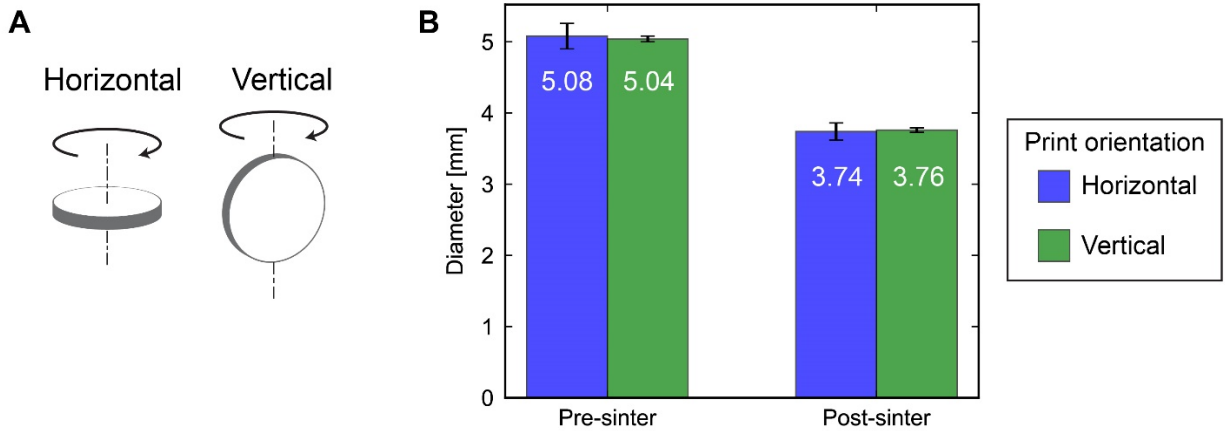

**Fig. S10. Shrinkage shows no dependence on print orientation.** (**A**) Diagram depicting the orientation in which the disks were printed with respect to the rotation axis of the print container. (**B**) Sintering results in about a linear shrinkage of about 26% (by the measured decrease in diameter) for both horizontally and vertically printed disks. Samples were printed, rinsed, and thermally debinded and subsequently their diameters were measured optically. After sintering their diameters were measured again. Number of samples: $n_{horz}$ = 11 and $n_{vert}$ = 9. Standard deviation: $\sigma_{pre\text{-}sinter,horz}$ = 0.18 mm, $\sigma_{pre\text{-}sinter,vert}$ = 0.04 mm, $\sigma_{post\text{-}sinter,horz}$ = 0.12 mm, $\sigma_{post\text{-}sinter,vert}$ = 0.03 mm.

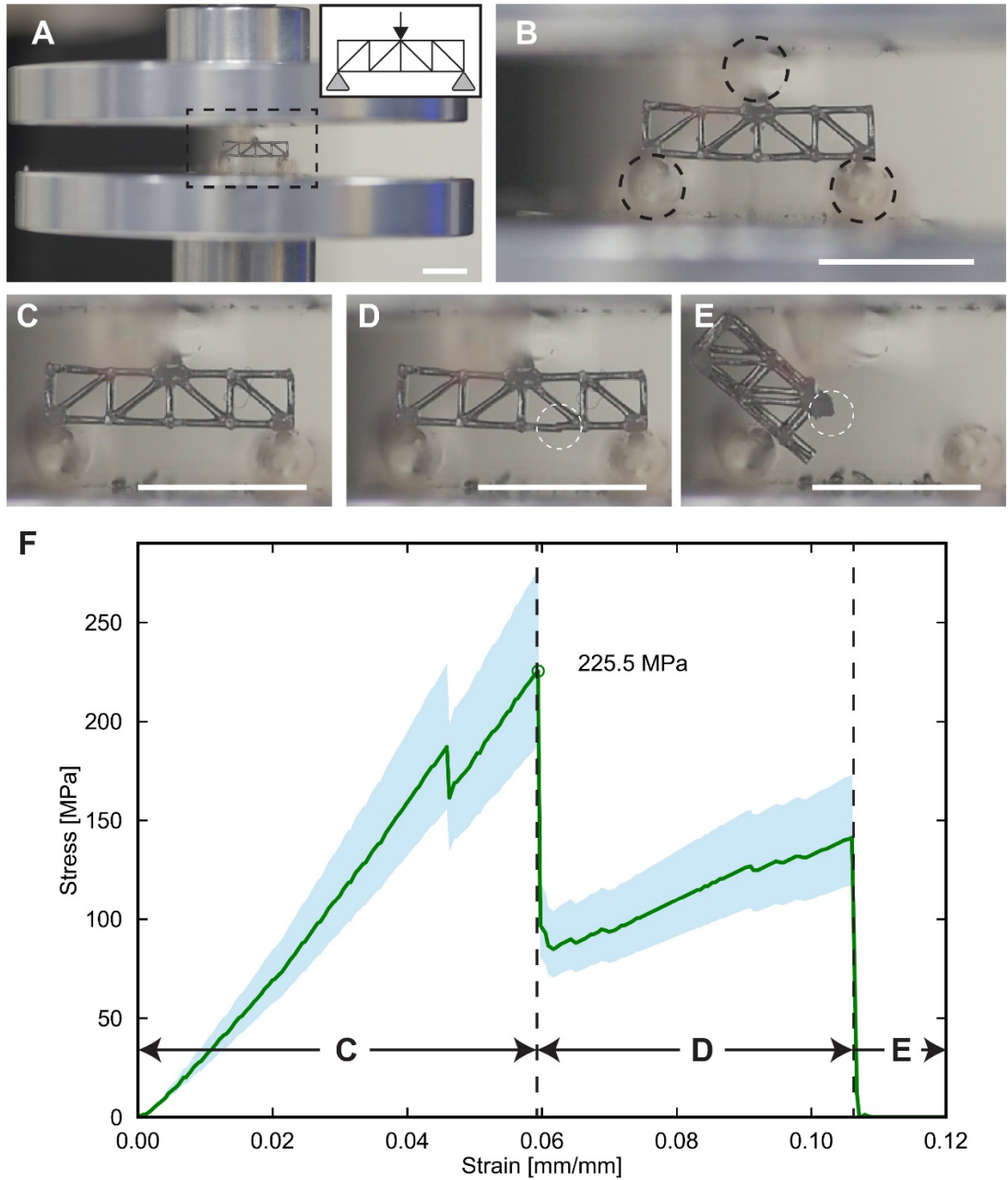

**Fig. S11. Mechanical testing of Howe truss.** (**A**) Compressive loading plates with Howe truss sample fixtured. (**B**) Zoomed image showing the cylindrical supports and "anvil". Note: the ends of the cylinders appear blurred in the image because they are outside of the focal region of the camera's lens. Image of: (**C**) linear elastic loading, (**D**) failure of bottom strut indicated by dashed circle, and (**E**) complete failure of truss with failure location indicated by dashed circle. (**F**) Tensile stress in the member indicated in D vs strain where regions C–E refer to the physical states shown in images C–E. The modulus of rupture is 225.5 MPa. Scale bars: 5 mm.

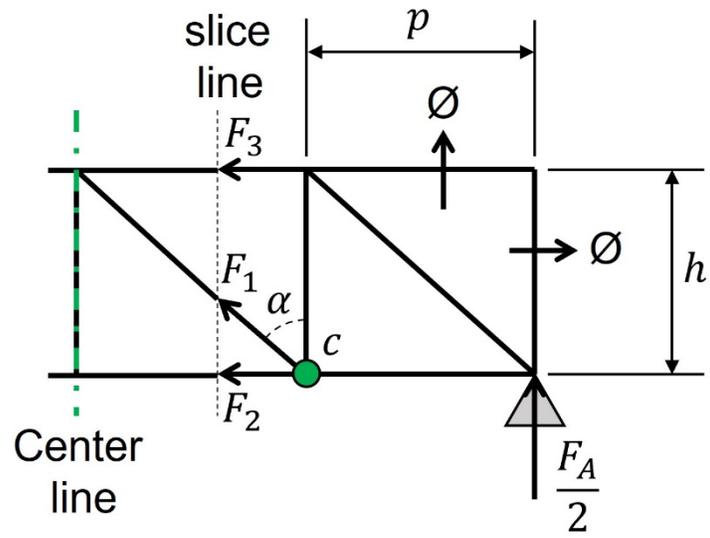

**Fig. S12.** Free-body diagram of the Howe truss under 3-point bend loading.

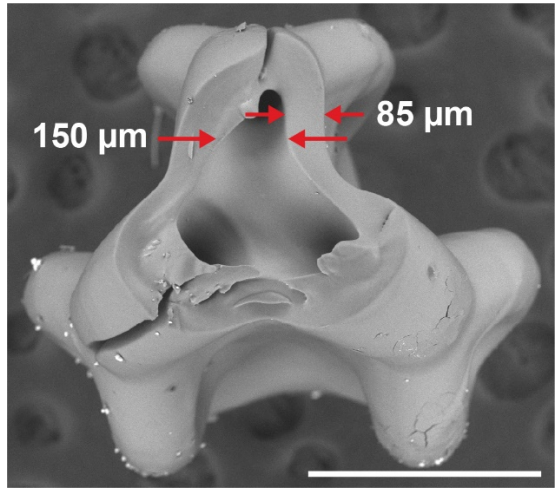

**Fig. S13. Sectioned branched 3D microfluidic device.** Minimum channel diameter of 150 µm and wall thickness of 85 µm are indicated in the SEM micrograph. Scale bar: 500 µm.

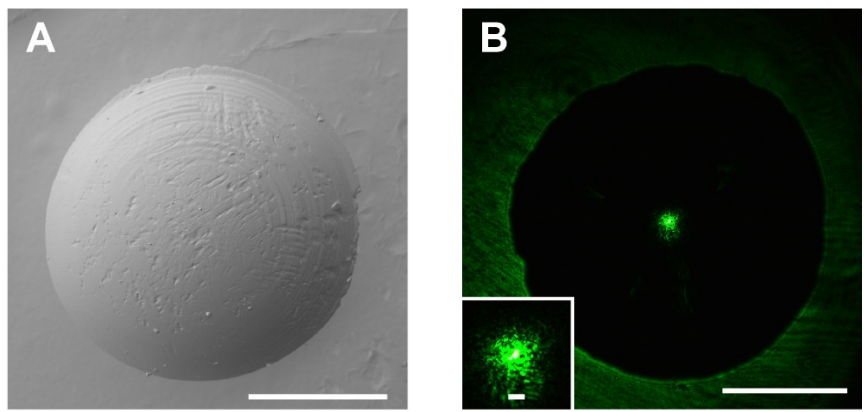

**Fig. S14. CAL-printed sintered planoconvex spherical lens.** (**A**) Topographical SEM micrograph of the sintered convex surface. Scale bar: 1 mm. (**B**) PSF of the lens under collimated 532 nm illumination. Scale bars: 1 mm (inset: 50 µm).

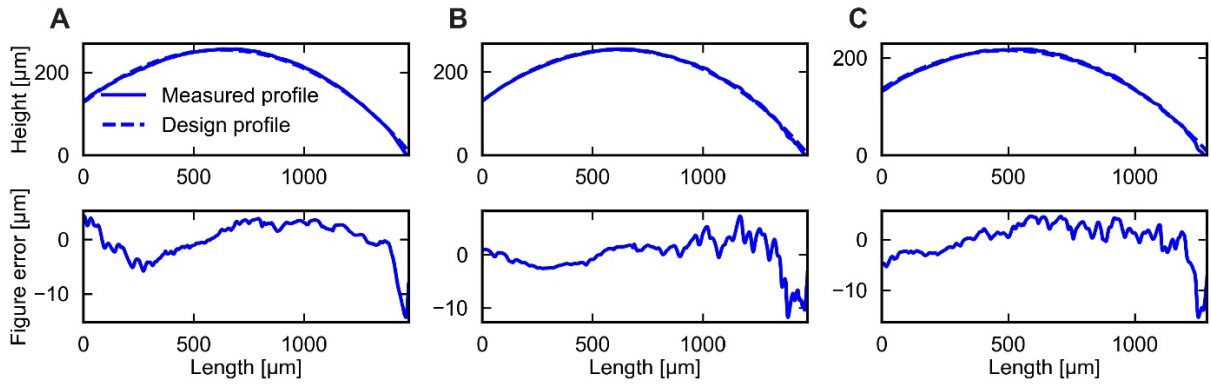

**Fig. S15. 1D figure error of the convex surface of the spherical lens in Fig. S14**. (**A–C**) Three different samples were measured with contact profilometry (Dektak Veeco 8 stylus profilometer; 12.5 µm radius stylus). The upper row shows the measured profile compared to an arc with the design radius of curvature of 1.6 mm. The constrained-radius arc was fit to the data using non-linear least squares optimization. The lower row shows the figure error between the measured and design profiles.

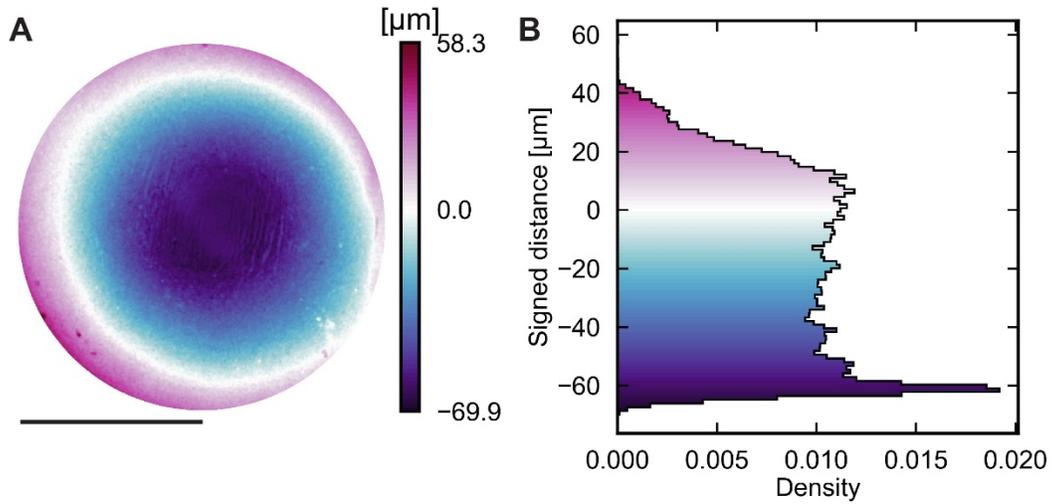

**Fig. S16. As-printed lens figure error.** The 3D profile of the as-printed top lens surface in Fig. 4F is measured with laser scanning confocal profilometry (Keyence VK-X1000). (**A**) Pseudo-color form error quantified as the signed distance from the printed surface to the reference design surface. Negative distance indicates the printed surface is eroded relative to the reference and positive distance indicates the printed surface is dilated relative to the reference. Scale bar: 1 mm. (**B**) Histogram of the signed distance. The color corresponds to the pseudo-color plot in A.

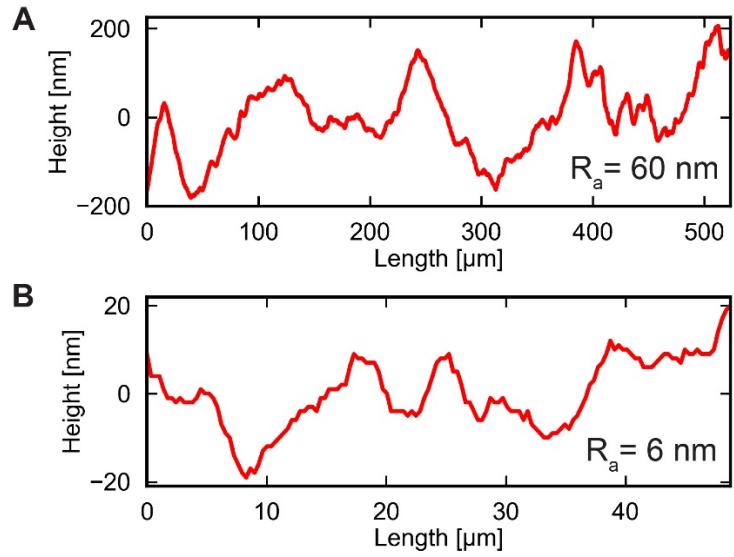

**Fig. S17. Sintered roughness.** Roughness of the sintered top lens surface shown in Fig. 4F is measured with laser scanning confocal profilometry. (**A**) Profile measured over 523 μm with $R_a$ = 60 nm. (**B**) Profile measured over 49 μm with $R_a$ = 6 nm.

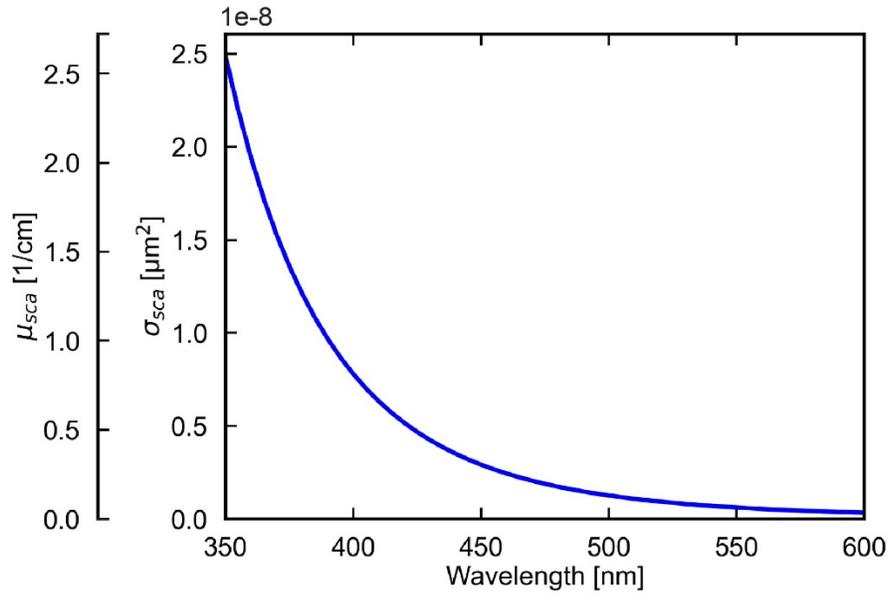

**Fig. S18. Simulated scattering cross section and scattering coefficient vs wavelength.** The scattering coefficient is calculated with $r$ = 0.02 μm, $N$ = 1.0444×10$^4$ μm$^{-3}$, and $\phi$ = 0.35.

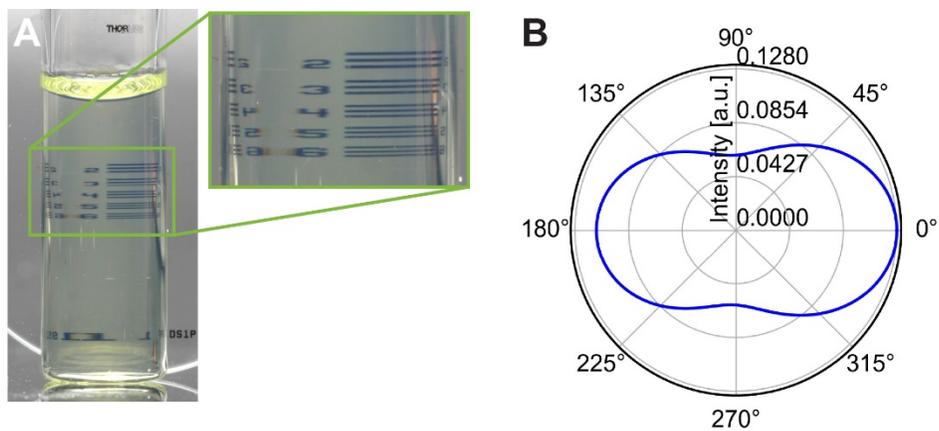

**Fig. S19. Resin light scattering.** (**A**) µSL v2.0 resin in a shell vial in front of a USAF resolution target. The outer diameter of the vial is 8 mm. (**B**) Scattering phase function of µSL v2.0 nanocomposite at 442 nm.

**Table S1. Optomechanical components**

| Component identifier in Fig. 1A in main text | Component |
|---|---|
| L | 3 W 442 nm laser module |
| L1 | Thorlabs RMS20X |
| SF | Mitsubishi Cable Industries CP95P2 ST50QE square core fiber; NA = 0.2, 50 × 50 μm$^2$ |
| L2 | Thorlabs A397TM-A |
| M1, M2, M3 | Thorlabs BB1-E02 |
| L3 | Thorlabs LC1054-A |
| L4 | Thorlabs LA1433-A |
| DMD | Vialux V-9501 |
| L5 | Thorlabs LA1509-A |
| L6 | Thorlabs RMS4X (0.45× config); Thorlabs LA1986-A (1.25× config); |
| A | Thorlabs SM1D12 |
| V | ThermoFisher C4015-96 |
| LED | Red 5 mm LED |
| L7 | Thorlabs AC254-030-A |
| L8 | Thorlabs RMS4X |
| L9 | Thorlabs LA1257-A |
| CCD | Amscope MU800 |
| Rotation stage | Thorlabs PRM1Z8 |

**Table S2.** Cauchy fit parameters $n(\lambda) = A + \frac{B}{\lambda^2} + \frac{C}{\lambda^4}$

| Parameter | Binder | µSL v2.0 |
|---|---|---|
| A | 1.45112863 | 1.44827419 |
| B | 0.00480647559 | 0.00520207580 |
| C | 0.000133597110 | -0.0000395412459 |

**Table S3. Debinding protocol**

| Debinding step | Target temperature (°C) | Heating rate (°C/min) | Dwell time (min) |
|---|---|---|---|
| 1 | 150 | 0.5 | 120 |
| 2 | 320 | 0.5 | 240 |
| 3 | 600 | 0.5 | 240 |
| 4 | 25 | 10 | - |
|  |  | **Total time:** | 30.1 hours |

**Table S4. Sintering protocol**

| Sintering step | Target temperature (°C) | Heating rate (°C/min) | Dwell time (min) |
|---|---|---|---|
| 1 | 1300 | 3 | 180 |
| 2 | 25 | 3 | - |
|  |  | **Total time:** | 17.2 hours |

**Table S5. Dimensions of sintered Howe trusses**

| Dimension | Sample 1 (Fig. 4A and B) | Sample 2 (Fig. S11) |
|---|---|---|
| $p$ | 1.66 mm | 1.59 mm |
| $h$ | 1.35 mm | 1.33 mm |
| $\alpha$ | 50° | 50° |
| $r_{mean}$ | 305 µm | 220 µm |
| $r_{min}$ | 290 µm | 190 µm |
| $r_{max}$ | 330 µm | 250 µm |

**Table S6. Truss members profilometry (Keyence VK-X1000)**

| Member | $R_a$ (µm) | $R_z$ (µm) | Measurement length (µm) |
|---|---|---|---|
| 1 | 0.367 | 1.836 | 135 |
| 2 | 0.223 | 1.528 | 180 |
| 3 | 0.716 | 3.020 | 148 |
| 4 | 0.298 | 1.821 | 219 |
| 5 | 0.331 | 2.558 | 218 |
| 6 | 0.513 | 2.235 | 165 |
| 7 | 0.446 | 3.153 | 191 |
| 8 | 0.492 | 2.274 | 172 |
| 9 | 0.393 | 1.960 | 175 |
| 10 | 0.321 | 1.251 | 205 |
| 11 | 0.377 | 2.004 | 128 |
| Mean | 0.407 | 2.149 | 176 |
| Std. dev. | 0.127 | 0.557 | 29.5 |